\documentclass[twocolumn,showpacs,preprintnumbers,pre]{revtex4}
\usepackage{graphicx}
\usepackage{dcolumn}
\usepackage{bm}
\usepackage{textcomp}
\usepackage{amssymb}
\usepackage{amsmath}

\newcommand{\be}{\begin{equation}}   \newcommand{\ee}{\end{equation}}
\newcommand{\ba}{\begin{array}}      \newcommand{\ea}{\end{array}}
\newcommand{\bea}{\begin{eqnarray}}  \newcommand{\eea}{\end{eqnarray}}

\begin{document}

\preprint{}

\title{Velocity Inversion in Nanochannel Flow}

\author{Youngkyun Jung}
\affiliation{Supercomputing Center, Korea Institute
of Science and Technology Information, P.O. Box 122, Yuseong-gu,
Daejeon 305-806, Korea}


\begin{abstract}
The nanoscale cylindrical Couette flow is investigated by means of
molecular dynamics simulations, in the case where the inner cylinder
is rotating whereas the outer cylinder is at rest. We find that the
tangential velocity of the flow is inverted when the fluid-wall
interaction near the outer cylinder is weak and the fluid density is
low. The unusual velocity inversion behavior is shown to be strongly
related to the degree of the slip between the fluid and the outer
cylinder, which is determined by the presence or absence of the
layering of the fluid near the outer wall.
\end{abstract}


\maketitle

\section{Introduction}
Couette flow between two concentric rotating cylinders shows
interesting and unexpected feature such as ``velocity inversion"
which implies that the tangential velocity of the flow increases
with distance from a rotating cylinder to a stationary cylinder. The
velocity inversion phenomenon has been studied with analytical and
numerical methods. Einzel {\it et al.}\cite{einzel90} first
predicted by suggesting a generalized slip boundary condition for
incompressible flow over curved or rough surfaces that the velocity
profile would become inverted in the case of large velocity slip at
the wall surfaces. Tibbs {\it et al.}\cite{tibbs97} and Aoki {\it et
al.}\cite{aoki03} confirmed the prediction of the anomalous behavior
using direct simulation Monte Carlo calculations and some analytic
approaches and found that the values of the accommodation
coefficients in their models play an important role on the velocity
inversion phenomenon. In all those studies, they found that the
velocity inversion phenomenon only occurred for large velocity slip
at small value of the accommodation coefficient. The accommodation
coefficient is an important parameter in determining the degree of
slip of the fluid at the wall and represents the average tangential
momentum exchange between the flowing fluid particles and the wall
boundary. The case of zero value of the coefficient is called
specular reflection, meaning zero friction. When its value is unity,
the reflection is diffuse, meaning that the fluid particles are
reflected with zero average tangential velocity. Recently, similar
studies for the velocity inversion phenomenon were performed
\cite{lockerby04,yuhong05,myong05}. It, however, is difficult that
the concept of an accommodation coefficient can apply to all types
of fluids or the nature of the wall materials.

Molecular Dynamics (MD) is a very powerful tool in the exploration
and study of the nature of the flow at the boundaries independent of
the properties of the fluid and wall materials. For simple
Lennard-Jones liquids, the wetting properties of the fluid can be
modeled by wetting parameter varying the strength of the fluid-wall
attraction\cite{barrat99,koplik,koplik01}. The wetting parameter
could directly be related to the accommodation coefficient of the
fluid-wall system. When the value of parameter is zero (nonwetting
or specular reflective), the fluid-wall interface is specular
reflective similar to the case of zero value of the accommodation
coefficient. When it is unity (wetting or attractive), the surface
is diffuse reflection. The large slip at the fluid-wall interface
exhibits under nonwetting condition and it decreases as the value of
wetting parameter increases to unity. The degree of the slip is also
dependant on the density of the fluid. Even at a strongly attractive
wall, a large slip was found in the highly dilute gas
regime~\cite{koplik01}.

In this paper we use MD simulations for Lennard-Jones liquids to
investigate the velocity inversion phenomenon in nanoscale
cylindrical Couette flows of the concentric rotating cylinders,
varying the fluid density and the value of the wetting parameter.
The tangential velocity of the flow is measured in the case where
the inner cylinder is rotating whereas the outer cylinder is at
rest. At low fluid density and under the poor wetting condition, the
velocity is inverted with large velocity slip near the outer
cylinder. On the other hand, at high density or under the good
wetting condition, such inversion does not occur with the layer
adsorption of the fluid near the outer cylinder. Next, the radial
force field is considered, and it is found that the radial force is
a better measurement than the fluid density to examine the effect of
the layering near the outer cylinder on the velocity slip.

\begin{figure}[bbp]
\includegraphics[width=8.0cm,angle=0]{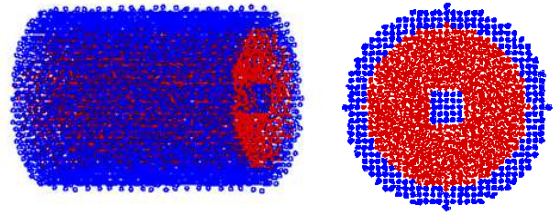}
\caption{(Color online) Views of a fluid in a nanochannel
consisting of two concentric cylinders.} \label{fig:cylinder}
\end{figure}

\section{Molecular Dynamics Simulations}
To investigate the properties of the nanochannel flows in a curved
surface, we have conducted MD simulations of the cylindrical
Couette flows in the concentric rotating cylinders as shown in
Fig.~{\ref{fig:cylinder}}. Standard MD techniques are
used\cite{allen,koplik05}. The fluid particles and the wall
particles at a distance $r$ interact through a Lennard-Jones(LJ)
potential,
$V_{ij}(r)=4\epsilon[(r/\sigma)^{-12}-A_{ij}(r/\sigma)^{-6}]$,
where $\epsilon$ and $\sigma$ represent the energy and length
scales, and $A_{ij}=A_{ji}$ is a dimensionless parameter that
controls the attractive part of the potential for the fluid-fluid
and the fluid-wall particles. As mentioned previously, this
parameter is similar to the accommodation coefficient in continuum
descriptions\cite{einzel90,tibbs97,aoki03,lockerby04,yuhong05,myong05}.
The potential is truncated at $r_{c}=2.5\sigma$. Newton's
equations are integrated with a velocity Verlet
algorithm\cite{allen} with a time step $\delta t=0.005\tau$, where
$\tau=\sigma (m/\epsilon)^{1/2}$ represents the characteristic
time scale with fluid particle mass $m$. A Dissipative Particle
Dynamics (DPD) thermostat\cite{soddemann} is used to keep the
system at a constant temperature $T=1.0 \epsilon/k_{B}$, where
$k_{B}$ is the Boltzmann constant. A coupling constant and a
weight function are chosen to be $\zeta=0.5 \tau^{-1}$ and
$w(r_{ij})=1-\frac{r_{ij}}{r_{c}}$, respectively, with the same
cutoff $r_{c}=2.5 \sigma$ as the LJ potential (for further details
see Ref.\cite{soddemann}). A Langevin thermostat\cite{grest} is
also considered to check the correctness of our results. The
thermostat with damping constant $0.1\tau^{-1}$ is only applied in
the axial direction of the cylinder in which the fluid is not
being sheared. We have checked that both thermostats give the
consistent results confirming the appearance of the velocity
inversion phenomenon in our system. Thus, in this paper, we only
present results obtained with the DPD thermostat.

The fluids are confined to a gap between two concentric cylinders,
whose walls are composed of particles of mass $m_{w}=100m$,
tethered by a stiff linear spring with constant $k_{w}=100$ to
fixed lattice sites. The fluid particles interact via LJ potential
with the $A_{ff}=1$. The channel wall are made of the same
material with the particle and interact with fluid particle with
$A_{fw}=1$ at the inner cylinder and $A_{fw}=0$ at the outer
cylinder. Periodic boundary condition is used in the axial
direction. All simulations are performed with a fixed inner
cylinder radius as $R_{i}/\sigma= 2.05$ (4 molecular diameters)
and different radius of the outer cylinder ranging from $10.26$
(20 molecular diameters) to $27.36$ (53 molecular diameters).

\begin{figure}[ttp]
\includegraphics[width=6.0cm,angle=270]{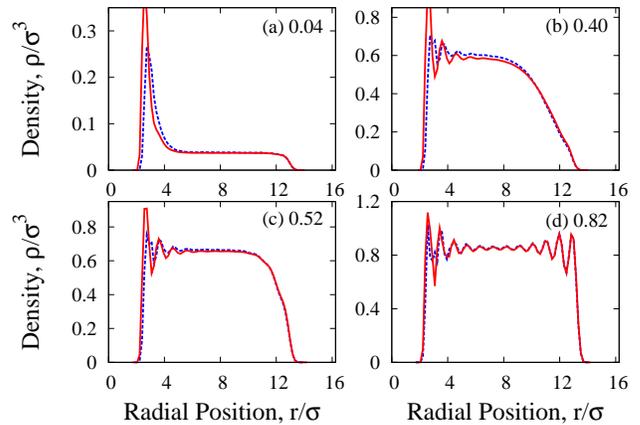}
\caption{(Color online) Density profiles for the angular
velocities $\omega=0.0\, \mbox{rad}/\tau$ (blue dotted line) and
$\omega=0.1\, \mbox{rad}/\tau$ (red solid line) with different
fluid densities $0.04, 0.40, 0.52$, and $0.82$.} \label{fig:den}
\end{figure}

\begin{figure}[bbp]
\includegraphics[width=6.0cm,angle=270]{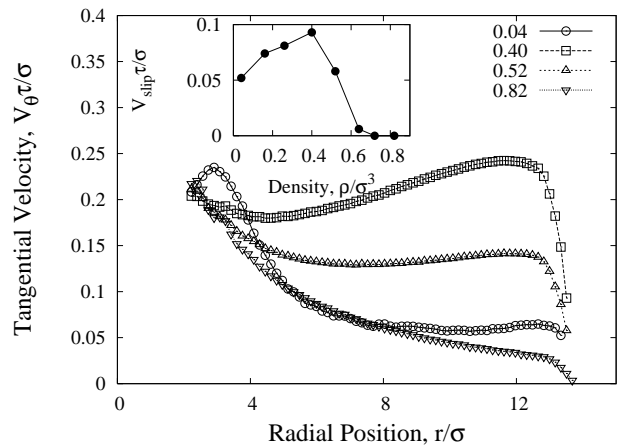}
\caption{Tangential velocities of the flows as a function of the
radial positions at different volume-average number densities of
the fluids. The inset shows the slip velocities as a function of
the fluid density, which are measured at the closest bin to the
outer cylinder.} \label{fig:den_vel}
\end{figure}

\section{Numerical Results and Discussion}
\subsection{Density and tangential velocity profiles}
The fluid is sheared by rotating the inner cylinder at a constant
angular velocity $\omega=0.1\, \mbox{rad}/\tau$, while the outer
cylinder remain stationary for all the simulations. After an
equilibration time of $5 \times 10^4 \tau$, the averages are
obtained by dividing the interior of the tube into cylindrical
shells of thickness $\sigma/4$ for a period of about $2 \times
10^5 \tau$. Figure~\ref{fig:den} shows the density profiles for
both the stationary ($\omega=0.0\, \mbox{rad}/\tau$) and the
rotating ($\omega=0.1\, \mbox{rad}/\tau$) inner cylinder denoted
by the blue dashed and the red solid line, respectively. The fluid
densities used in Fig.~\ref{fig:den} which are volume-average
number densities of the fluids, $\rho/\sigma^{3}$, are $0.04,
0.40, 0.52$, and $0.82$. There are fluid layers near the inner
cylinder corresponding to density peaks as a consequence of the
attractive interaction ($A_{fw}=1$) between the fluid particles
and the inner cylinder, while near the outer cylinder with
$A_{fw}=0$ the peak does not exist except for dense regime (see
Fig.~\ref{fig:den}(d)). For dense regime, however, a large number
of layers are observed near the outer cylinder due to the highly
packing of the particles. When the inner cylinder is rotated, the
angular momentum attained from the rotating cylinder is
continuously transferred to the outer region through the
inter-particle interactions, so that the particles near the inner
cylinder are pushed out to the outer region.

We have carried out the simulations at different fluid densities
ranging from $0.04$ to $0.82$ to measure the tangential velocity
of the flows. The tangential velocity initially decreases with the
radial position due to the momentum dissipation through
inter-particle interactions as shown in Fig.~\ref{fig:den_vel}.
It, however, is shown that the velocities are inverted at below a
certain fluid density about $\rho/\sigma^3 \approx 0.6$. In other
words, the velocity increases with distance from the rotating
inner cylinder. This unusual behavior of the tangential velocity
is similar to the results of the previous
studies\cite{einzel90,tibbs97,aoki03,lockerby04,yuhong05}. The
velocities finally become decaying again near the outer cylinder,
which will be explained in detail below. It is well known that the
slip velocity at the stationary outer cylinder plays crucial role
of the velocity inversion\cite{einzel90}. In our simulations such
slip velocity appear at below a certain fluid density
$\rho/\sigma^3 \approx 0.6$, below which the velocities are
inverted. Above $0.6$, no velocity slips and no tangential
velocity is inverted even though the fluid-wall interaction is
specular ($A_{fw}=0$). Inset of Fig.~\ref{fig:den_vel} shows the
slip velocity near the outer cylinder as a function of the fluid
density. The maximum value of the slip velocity is located at
$\rho/\sigma^3 =0.40$, where the degree of the velocity inversion
is largest. For low density, the slip velocity increases with
fluid density where the interactions of the particles with the
wall and with each other are dominant. However, for dense cases,
it decreases where viscosity becomes dominant, so that the flow
gets slow down. It is sure that the velocity inversions are shown
if the slip velocity is higher than a certain value.

\begin{figure}[tbp]
\includegraphics[width=6.0cm,angle=270]{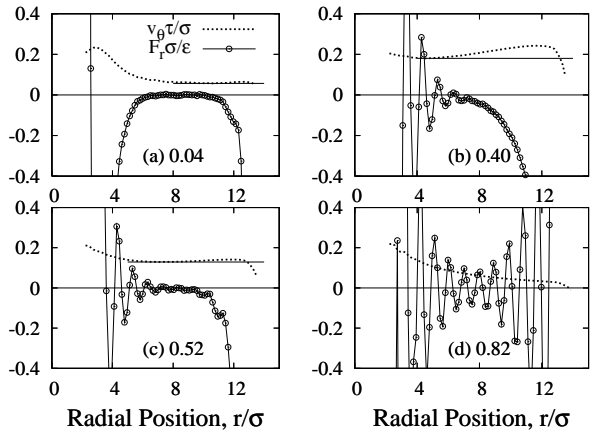}
\caption{Radial force profiles of the flow (circle) for the
indicated values of the fluid density $\rho/\sigma^{3}$ ($0.04,
0.40, 0.52$ and $0.82$) as a function of the radial position at
$\omega=0.1\, \mbox{rad}/\tau$. The tangential velocity (dotted
line) profiles in Fig.~\ref{fig:den_vel} are added to compare them
with the radial force profiles. The horizontal solid lines are
used as guidelines to the eye.} \label{fig:den_vel_fr}
\end{figure}

\begin{figure}[tbp]
\includegraphics[width=6.0cm,angle=270]{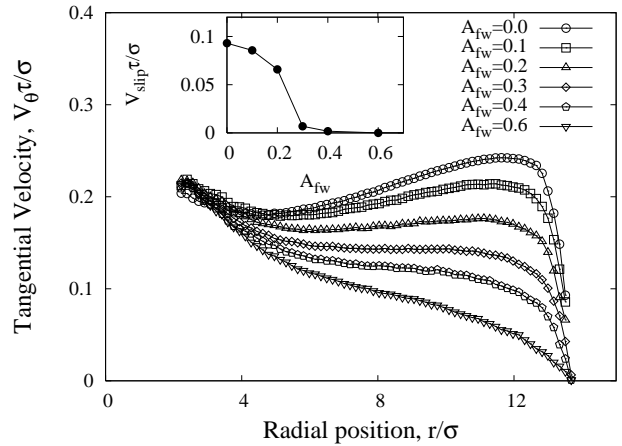}
\caption{Tangential velocity profiles with different $A_{fw}$
ranging from $0.0$ to $0.6$. The fluid density is fixed at
$\rho/\sigma^{3}=0.40$. The inset shows the slip velocity near the
stationary outer cylinder as a function of $A_{fw}$.}
\label{fig:vel_A}
\end{figure}

\subsection{Radial force profile}
In order to gain an insight on such an unusual inverted velocity
profile in the case of a stationary outer cylinder and a rotating
inner cylinder, we measure the radial force generated by both the
rotating inner cylinder with $A_{fw}=1$ and the stationary outer
cylinder with $A_{fw}=0$. Figure~\ref{fig:den_vel_fr} shows the
radial force profiles at different fluid densities, corresponding
to $0.04$, $0.40$, $0.52$, and $0.82$. The tangential velocity
profiles in Fig.~\ref{fig:den_vel} are added to compare them with
the radial force profiles. It is important to put our finger on
the appearance of the peaks in the radial force profiles, as in
the density profiles (see Fig.~\ref{fig:den}). However, compared
to the peaks of the density, the oscillations of the radial force
are shown more clearly. In the vicinity of the inner cylinder,
oscillations of the radial force reveal due to the adsorption of
the fluid at the cylinder. On the other hand, near the reflective
outer cylinder there exist no such oscillation except for the
dense regime as in Fig.~\ref{fig:den_vel_fr}(d). In
Figs.~\ref{fig:den_vel_fr}(a), (b), and (c), there is a point at
which the outward motion of the fluid meets to the inward motion,
which results from the competition between the centrifugal force
by the rotating inner cylinder and the centripetal force by the
reflective outer cylinder. Moreover, due to the lack of fluid-wall
interaction near the outer cylinder, the angular momentum attained
from the rotating inner cylinder is well transferred to the outer
region and induces the velocity slip. It is interesting to note
that the tangential velocity begin to be inverted at the point.
However, there is no such a point for dense case as shown in
Fig.~\ref{fig:den_vel_fr}(d). In this regime, the fluid-wall
interaction near the outer cylinder are very strong, so that the
fluids in the first layer can hardly slip and the first fluid
layer induces a second layer by the fluid-fluid interaction. The
second layer induces a third, and so on. Based on these
observations, it is sure that the presence of the layering in the
vicinity of the outer cylinder is a key factor to determine
whether the velocity inversion phenomenon appears or not.

\begin{figure}[tbp]
\includegraphics[width=7.0cm,angle=270]{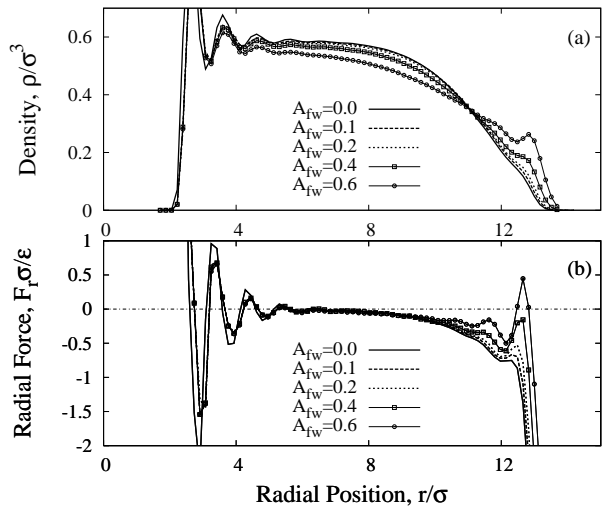}
\caption{Density (a) and radial force(b) profiles of the flows as
a function of the radial position at $\omega=0.1\,
\mbox{rad}/\tau$ with several values of $A_{fw}$. The fluid
density is fixed at $\rho/\sigma^{3}=0.40$. The dotted line in
Fig. (b) denotes a zero value of the radial force at which the
sign of the radial force is changed.} \label{fig:A_den_fr}
\end{figure}

\subsection{Layering effect}
To understand the effect of the layering near the outer cylinder
on the velocity inversion phenomenon, we measure the tangential
velocity with different values of interaction parameter $A_{fw}$
between the fluid and the outer cylinder at fixed fluid density
$\rho/\sigma^{3}=0.4$. At this density, the velocity inversion
behavior can be seen most clearly (see Fig.~\ref{fig:den_vel}). As
$A_{fw}$ increases, the degree of the velocity inversion gradually
decreases and finally disappears at exceeding $A_{fw}=0.3$ as
shown in Fig.~\ref{fig:vel_A}. The slip velocity shown in the
inset of Fig.~\ref{fig:vel_A} for six independent realizations of
different $A_{fw}$, corresponding to $0.0, 0.1, 0.2, 0.3, 0.4,$
and $0.6$ also decreases to zero. These indicate that the velocity
inversion behavior is strongly related to the strength of the
fluid-wall interaction near the outer cylinder. The responses of
the density and the radial force with increasing the value of
$A_{fw}$ is shown in Fig.~\ref{fig:A_den_fr}. As the value of
$A_{fw}$ increases, the outer cylinder becomes more attractive so
that more fluid particles can move to the outer region and the
adsorbed layers begin to form near the outer cylinder. This
formation of the adsorbed layers can be seen most clearly in the
radial force profile rather than in the density profile as
revealed in Fig.~\ref{fig:A_den_fr}. Even for $A_{fw}=0.0$, an
indication of a peak of the radial force is shown. The fluid
particles of the first layer almost stick to the wall, which
causes the flow to retard in the vicinity of the wall. The rapid
decay of the tangential velocity near the outer cylinder shown in
Fig.~\ref{fig:vel_A} can be explained by this layering effect even
though it is weak. When the value of $A_{fw}$ is small (less than
$0.3$), the first layer adsorption is only appeared weakly and its
size gradually increases as $A_{fw}$ increases from zero. For
small value of $A_{fw}$, the velocity of the flow can largely slip
and the velocity can be inverted. As $A_{fw}$ increases beyond
$0.3$, the effect of the first layer is getting strong and induces
a second layer, the second layer induces a third layer, and so on.
From the layer adsorption point of view, the fluid particles of a
second layer are retarded by the first layer, a third layer is
retarded by the second layer, and so on. As the number of fluid
layer increases, the no-slip nature at fluid-wall interface
spreads into the bulk, so that the flow can not slip and no
velocity inversion appears. It is finally concluded that the
appearance of the velocity inversion phenomenon is determined by
whether the layer adsorption near the outer cylinder is present or
not.

\begin{figure}[tbp]
\includegraphics[width=6.0cm,angle=270]{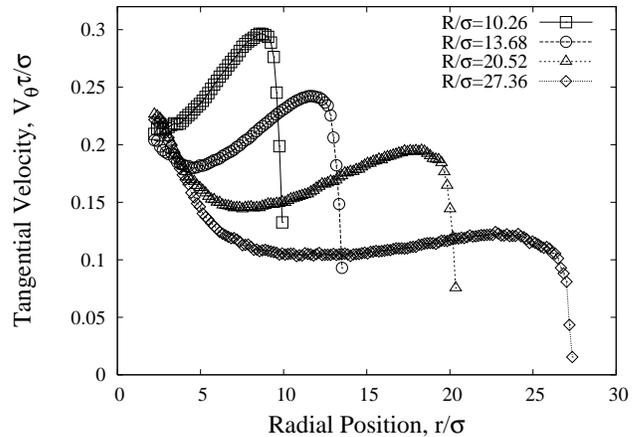}
\caption{Tangential velocity profiles with four different radius
of the outer cylinders at given fluid density
$\rho/\sigma^{3}=0.40$ and fixed angular velocity $\omega=0.1\,
\mbox{rad}/\tau$.} \label{fig:vel_size}
\end{figure}

The response of the velocity inversion with increasing the radius
of the outer cylinder ranging from $R/\sigma=10.26$ to $27.36$ at
fixed fluid density $0.40$ and angular velocity $0.1\,
\mbox{rad}/\tau$ is shown in Fig.~\ref{fig:vel_size}. At
$R/\sigma=10.26$, the tangential velocity is fully inverted. As
the radius of the outer cylinder increases, the effect of rotating
cylinder is reduced, so that finally the velocity would not slip
any more and not be inverted at given angular velocity. We find
also that no inverted velocity profile appears in the case where
the inner cylinder is stationary whereas the outer cylinder is
rotating.

\section{Conclusions}
Using molecular dynamics simulation, we study the behavior of the
nanoscale cylindrical Couette flow between two concentric rotating
cylinders. In simulations, both DPD and Langevin thermostats which
are suitable for out-of-equilibrium simulations are used to keep
the system at a constant temperature. When the inner cylinder
rotates and the outer cylinder is at rest, we find that the
tangential velocity is inverted with a large velocity slip at the
outer cylinder. The large slip at the outer cylinder, which plays
an important role in the appearance of the inversion, occurs at
low fluid density and weak fluid-wall interaction near the outer
cylinder. We also find that the appearance of the inversion
behavior is strongly related to the presence or absence of the
layering of the fluid near the outer cylinder. The formation of
the layers can be seen most clearly in the radial force profile.


\begin{acknowledgements}
The author would like to thank Jysoo Lee for useful discussions.
This work was supported in part by a research fund from IBM KOREA.
\end{acknowledgements}

\end{document}